\documentclass[%
 reprint,
prl,
]{revtex4-2}

\usepackage{amsmath,amssymb}
\usepackage{graphicx}% Include figure files
\usepackage{dcolumn}% Align table columns on decimal point
\usepackage{bm}% bold math
\usepackage{xcolor}
\usepackage{siunitx}
\usepackage{hyperref}% add hypertext capabilities
\begin{document}

\preprint{APS/123-QED}

\title{Vortex lattice states of bilayer electron-hole fluids in quantizing magnetic fields}
\author{Bo Zou}
\affiliation{Department of Physics, University of Texas at Austin, Austin, TX 78712}
\author{A.H. MacDonald}
\affiliation{Department of Physics, University of Texas at Austin, Austin, TX 78712}
\date{\today}
\begin{abstract}
We show that the ground state of a weakly charged two-dimensional electron-hole fluid
in a strong magnetic field is a broken translation symmetry state with interpenetrating 
lattices of localized vortices and antivortices in the electron-hole-pair field.  
The vortices and antivortices carry fractional charges of equal sign but
unequal magnitude and have a honeycomb lattice structure that contrasts with the 
triangular lattices of superconducting electron-electron-pair vortex lattices.
We predict that increasing charge density or weakening magnetic field drives
a vortex delocalization transition that would be 
signaled experimentally by an abrupt increase in counterflow transport resistance.
\end{abstract}
\maketitle

\begin{figure}[t]
    \centering
    \includegraphics[width=\linewidth]{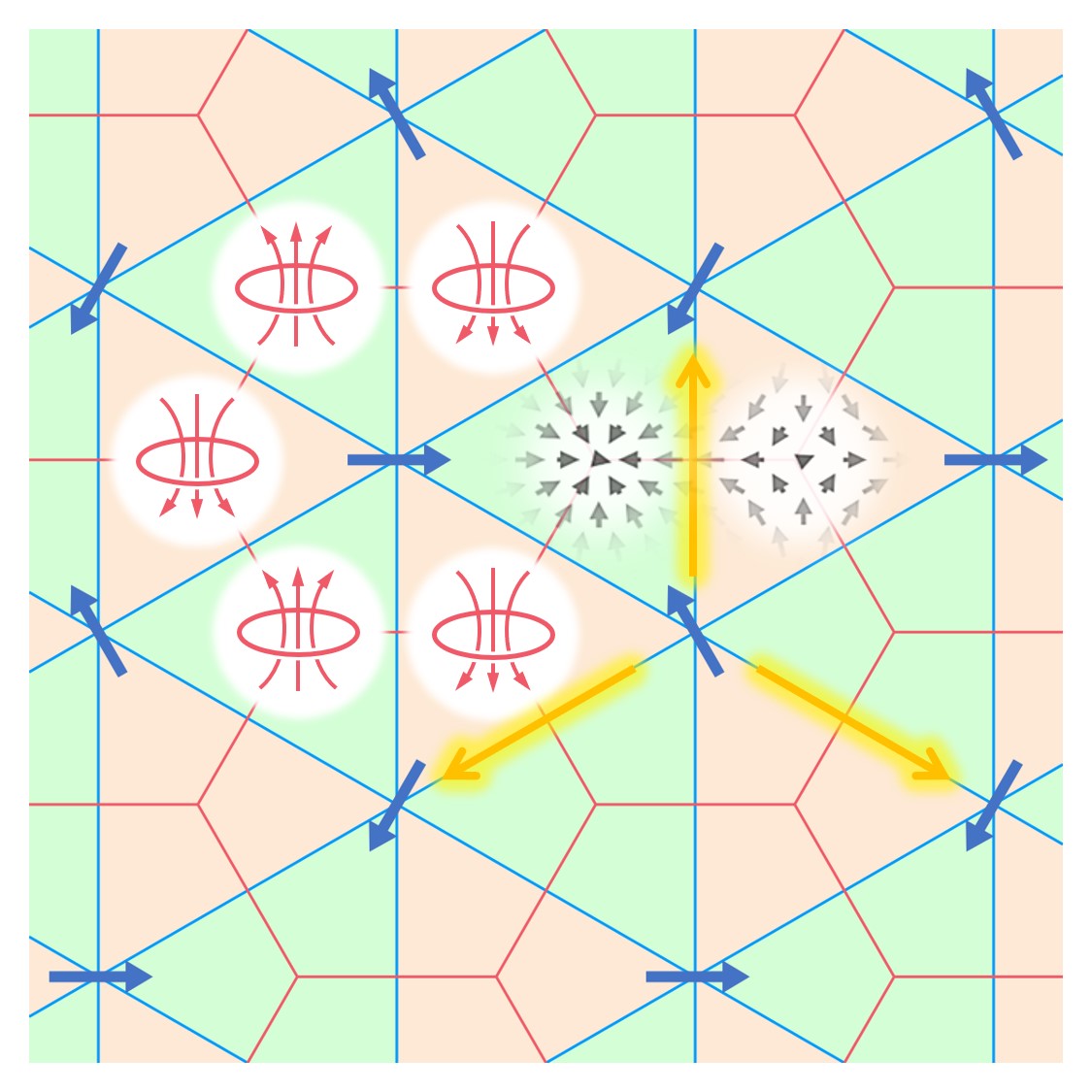}
    \caption{Schematic illustration of the vortex lattice mean-field ground states 
    of charged electron-hole fluids in a strong magnetic field. 
    The vortices and antivortices form a honeycomb lattice with red links. 
    The electron-hole pair amplitude vanishes at the vortex and antivortex centers 
    and peaks on the triangular lattice sites whose links are marked in blue.
    Each peak is marked by a blue arrow representing the local condensate phase.
    Phases differ by $\frac{2\pi}{3}$ across the near-neighbor links marked by gold arrows, 
    and by $-\frac{2\pi}{3}$ across links with opposite orientations.
    A Bose-Hubbard lattice model of this condensate contains alternating gauge fluxes in the green and orange triangles that originate from loop currents around vortices and antivortices.
    }
    \label{fig:latticeschematic}
\end{figure}

{\em Introduction.---}
Recent progress \cite{ ma2021strongly, gu2022dipolar, nguyen2025perfect, zeng2023exciton, qi2025perfect, qi2023thermodynamic, nguyen2025quantum, qi2025competition} in separately contacting 
electrons and holes located in electrically isolated but nearby 
two-dimensional (2D) semiconductor layers has 
opened up new opportunities to study quasi-equilibrium electron-hole systems with 
separately tunable \cite{xie2018electrical,zeng2020electrically} electron and hole densities.  
Because of the strong attractive 
interactions between electrons and holes, these systems have rich many-particle physics.
In previous work we have discussed how electron-hole correlations are strengthened in strong magnetic 
fields when electron and hole densities are equal \cite{zou2024electrical}, leading to robust electron-hole-pair condensates that exhibit \cite{nguyen2025quantum, qi2025competition} magnetic oscillations. 
Here we consider the case of a small but non-zero total charge density.
We find that, instead of a Wigner crystal of unit charge quasiparticles,
the ground state has a lattice of fractionally charged 
vortices and antivortices in the electron-hole-pair amplitude. 
% We find that vortices and antivortices in the electron-hole-pair amplitude
% host electric charge in the strong magnetic field case, 
% and demonstrate by explicit calculation that these charged order parameter textures have lower energy than the conventional electron or hole
% quasiparticles and are therefore present in the many-particle ground state. 

Vortex lattice states are common in systems with U(1) order parameters. 
Common examples include the vortex lattices in type-II superconductors \cite{abrikosov2004nobel} 
in external magnetic fields and in rotating atomic superfluids\cite{fetter2009rotating}.
In these systems, vortices are induced by magnetic fluxes and
rotations respectively and the total vorticity is proportional to system size.  
Because vortices and antivortices can annihilate, lattice
states with equal numbers of vortices and antivortices are less common,  
although they have been proposed as a theoretical possibility
in both $^4$He films and two-dimensional cold atoms \cite{zhang1993vortex,botelho2006vortex}, and can be generated by 
periodic external potentials\cite{hivet2014interaction,milovsevic2004vortex} in 
polariton fluids.  In lattice states with equal numbers of 
vortices and antivortices, the total vorticity vanishes.  In the 
electron-hole fluids we discuss, vortices and antivortices carry
electrical charges of the same sign so that repulsive Coulomb interactions 
disfavor annihilation.

\begin{figure*}
    \centering
    \includegraphics[width=0.95\linewidth]{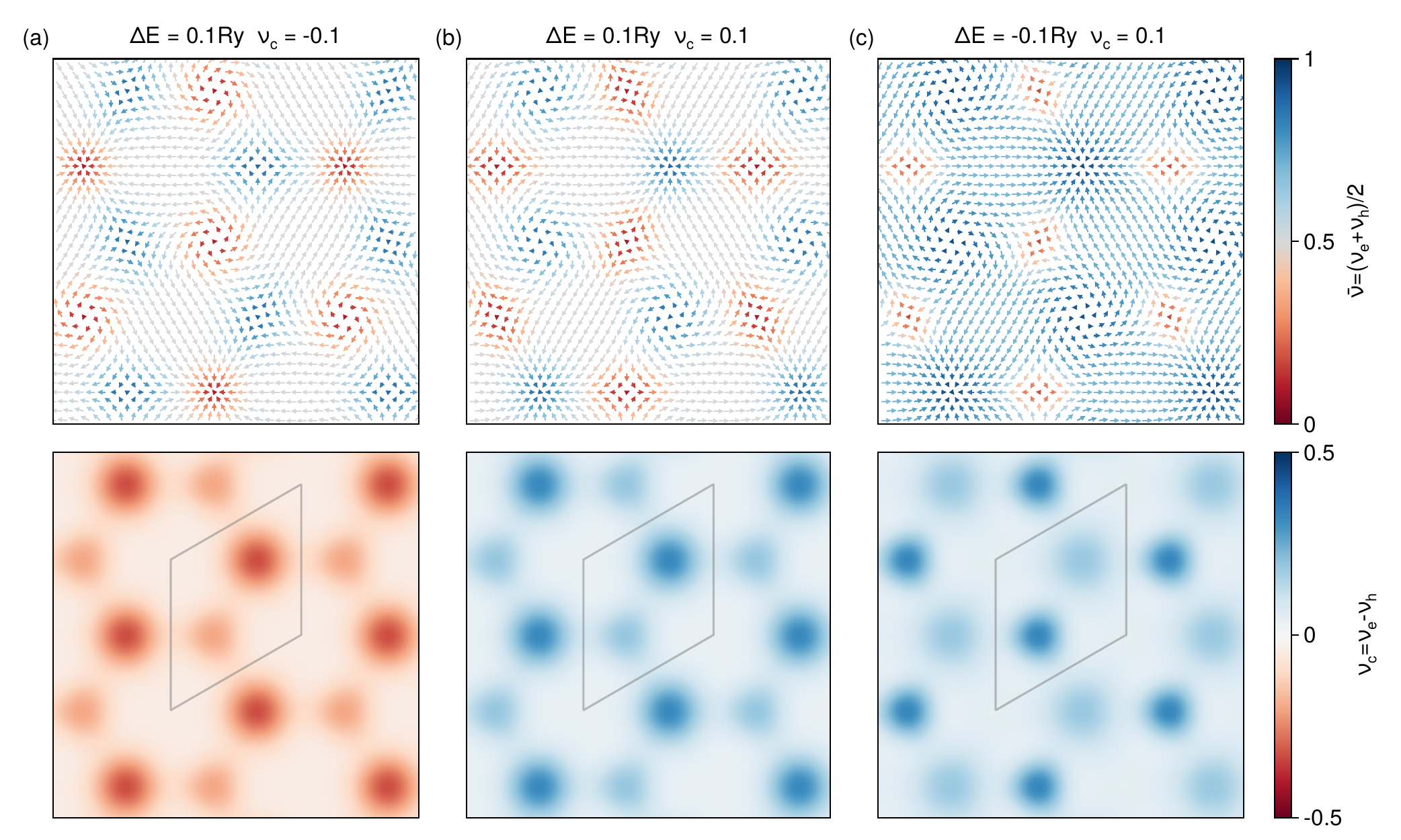}
    \caption{ 
    Honeycomb-lattice exciton vortex lattice states at magnetic field $B = 0.1B_0$ and 
    layer separation $d=a_B$ at different effective gap $\Delta E$ and averge charge filling factor $\nu_c$, 
    where $B_0$, $a_B$ and $Ry$ are the characteristic exciton magnetic field, length and energy scales defined in the main text. 
    The lengths and orientations of the arrows represent the electron-hole-pair
    condensate magnitudes and phases, respectively,
    and the color scales represent the average (top) and the difference (bottom) of the electron 
    and hole charge densities - both expressed as local Landau level filling factors.
    The comparison between (a) and (b) illustrates the particle-hole symmetry of our model whereas 
    the comparison between (b) and (c) illustrates the dependence on the effective band gap $\Delta E$.
    Four types of charged vortices, distinguished by vorticity and charge sign, 
    appear as generalizations of the half-charged (meron) vortices in balanced electron bilayers.
    The charge density distribution is sensitive to the effective band gap $\Delta E$, as 
    discussed in the main text.  
    These mean-field solutions were obtained at strong fields where Landau quantization is quite significant; results for weaker fields are illustrated in the supplementary material.
    The unit cell area (outlined in the bottom figures) of the broken translation symmetry state $A_{uc} = A_\Phi/|\nu_c|$, which includes one vortex and one antivortex, contains exactly one elementary charge.}
    \label{fig:real space}
\end{figure*}

In an electron-hole fluid at strong magnetic fields
the lowest energy charged excitations are vortex or antivortex textures in  
the electron-hole-pair order-parameter field.
They therefore appear in the ground state 
when a net charge density is induced by gate voltage settings.  
Unlike the charged Cooper pairs in superconductors, excitonic pairs do not accumulate 
Aharonov-Bohm phase by enclosing magnetic flux.  Any net 
vorticity per unit area therefore has a large gradient energy cost,
so that vortices and antivortices must have equal densities. 
The connection between charge-density and order-parameter texture 
closely parallels that of
Skyrmion charged spin textures \cite{sondhi1993skyrmions, brey1995skyrme} 
in quantum Hall ferromagnets and meron textures \cite{yang1994quantum, moon1995spontaneous, yang1996spontaneous} in electron-electron bilayers in the quantum Hall regime.
The charges hosted in the cores of a vortex and an antivortex sum to one elementary charge.
When electrons or holes are added to a neutral condensate, they fractionalize into 
vortex-antivortex pairs.  Our calculations show that once many of these charged textures are present 
they crystallize into the honeycomb lattice arrangement illustrated in Fig.~\ref{fig:latticeschematic}.

{\em Charged Electron-Hole Bilayers in Strong Perpendicular Magnetic Fields}---
We consider electron and hole layers that are separately contacted 
with voltages $V_{e,h}$ in the geometry discussed in 
Refs. \cite{xie2018electrical,zeng2020electrically, gu2022dipolar,zou2024electrical, ma2021strongly, qi2023thermodynamic} and in the supplementary material\cite{SM}.
When the leakage current between two layers is negligible, 
% where $l = (\hbar c/eB)^\frac{1}{2}$ is the magnetic length, 
the electrons and holes come to equilibrium with separate particle reservoirs and 
the system can be mapped to an equilibrium electron-hole system with 
separately conserved electron and hole numbers and an effective 
band gap $\Delta E =\epsilon_{c}-\epsilon_{v} +e(V_e - V_h)$.
Dual-gating, isolation between layers, and separate-contacting 
make it possible in combination to realize two-dimensional electron-hole fluids in which the  
total charge density and the effective band gap are separately electrically tunable,
and electron-hole recombination processes are absent -  
eliminating many of the subtle complications that obscure many-particle interaction physics
in optically pumped electron-hole systems.

We assume that a perpendicular magnetic field is present that is 
strong enough to permit truncation of both electron and hole Hilbert spaces to a finite 
number of of Landau levels.  For simplicity we also assume that both electrons and holes are 
fully spin-polarized by a combination of Zeeman and interaction effects, although this simplification 
is easily relaxed.  Choosing the middle of the effective gap as the zero of energy and 
taking electron and hole masses $m^*$ to be equal for definiteness,
the Landau level energies in the conduction and valence bands are 
$\epsilon_{n,c/v} = \pm \frac{1}{2} \Delta E \pm \left(n+\frac{1}{2} \right)\hbar\omega_c$.
% \begin{equation} \label{eq: Landau levels}
%     \left\{ 
%         \begin{aligned}
%             &\epsilon_{n,c} = \ \ \frac{1}{2} \Delta E + \left(n+\frac{1}{2} \right)\hbar\omega_c ,\\
%             &\epsilon_{n,v} = -\frac{1}{2} \Delta E - \left(n+\frac{1}{2}\right)\hbar\omega_c,\\
%         \end{aligned}
%     \right.
% \end{equation}
where $\omega_c=eB/m^*c$ is the cyclotron frequency.
Each Landau level has degeneracy $N_\Phi=A/A_\Phi$ with $A$ being the sample area and $A_\Phi=hc/eB$ being the flux quantum area, so that carrier densities $n$ and $p$ are related to Landau level filling factors by $n(p) = \nu_{e(h)}/A_\Phi$.  The total charge filling factor is defined as $\nu_c = \nu_e - \nu_h$.
% When the exciton binding energy exceeds the effective gap, bound electron-holes pairs are 
% present in the ground state.  

{\em Vortex Lattice States}---
We have previously discussed the strong field states of neutral electron-hole fluids 
($\nu_c$=0), focusing on the competition
between condensed and uncondensed phases induced by Landau kinetic energy quantization\cite{zou2024electrical}.
The uncondensed phases are integer quantum Hall phases in which both layers have fully occupied Landau levels. 
They appear only above critical magnetic field strengths beyond which large Landau level spacings
suppress exciton-binding coherence and give rise to unusual quantum oscillation behavior in insulating 
states that has now been confirmed experimentally \cite{nguyen2025quantum, qi2025competition}.
%States at fractional charge filling factors $\nu_c$ are expected to be either strongly correlated 
%fluid states that cannot be captured by the Hartree-Fock(HF) approximation we employ 
%or, at smaller $\nu_c$, the broken-translation-symmetry states on which we now focus.
To describe charged order-parameter-texture states,
we solve unrestricted Hartree-Fock equations that account for Landau-level mixing and 
broken translation symmetry using a convenient equation of motion approach  
that takes advantage of the analyticity of the Landau-level wavefunctions to 
express Green's functions, density matrices, and Fock potentials in terms of Fourier transforms of 
local quantities, greatly simplifying the calculations.
The technical details of these calculations are explained in the Supplementary Material \cite{SM}. 
We find that adding extra electrons or holes to the neutral exciton condensate 
leads to the honeycomb lattice 
\footnote{In addition to the honeycomb lattice solutions, we also find square vortex lattice solutions of the 
Hartree-Fock equations, but their energies are higher and we do not discuss them here.} of vortices and antivortices described above.
The Hartree-Fock equations also have uniform-density solutions in which the mean-field 
quasiparticles of the excitonic insulators accommodate the excess charge, but these have higher energy.

Fig.~\ref{fig:real space} illustrates the spatial variation of the electron and hole 
charge densities and the electron-hole-pair magnitude and phase for $d=a_B$ and $B=0.1B_0$.
(The exciton Bohr radius $a_B =2\epsilon\hbar^2/e^2m^*$, the Rydberg energy $Ry = e^2/2\epsilon a_B$, and the characteristic magnetic field 
scale $B_0$ is defined by $a_B^2B_0 =A_\Phi B$.
\footnote{For transition metal dichalcogenide (TMD) bilayers encapsulated by hexagonal
boron nitride (hBN), $a_B \approx 1.3$nm, $Ry\approx 0.11$eV, and $B_0 \approx 2.4\times10^3$T, whereas 
for GaAs quantum well systems, which have smaller masses and larger dielectric constants,
the corresponding scales are approximately $12$nm, $4.8$meV and $28$T.}.)
At this field strength electrons and holes occupy mainly the lowest few Landau levels. 
In Fig.~\ref{fig:real space}, the orientations and lengths of the arrows depict the electron-hole pair phase and magnitude and make the pattern of vortices and antivortices visible.  
The color scales indicate the sum and difference of the local 
electron and hole densities expressed as filling factors.  
The vortex lattice states in Figs.~\ref{fig:real space}(a) and (b) 
are at the same gap $\Delta E = 0.1Ry$ but have opposite $\nu_c=\pm0.1$.
Because of the particle-hole symmetry of the theory, the two results differ 
only in the sign of the charge density and in the vorticity of the vortices. 
In total, we recognize four types of vortices
\footnote{For brevity, we use vortices to refer to all the vortex and antivortex objects where  
it does not cause ambiguity.}
distinguished by their vorticity and 
fractional charge signatures, two of 
which appear on the electron side ($\nu_c >0$) and two on the hole side ($\nu_c<0$).
On each side, the two realized vortices have opposite 
vorticities and fractional charges that are unequal in magnitude but alike in sign. 
In the lattice, the charged vortices contribute in combination
one elementary charge per unit cell and have opposite layer polarizations in their core regions.
The unit cell area is therefore $A_{uc} = A_\Phi / |\nu_c|$. 
Each vortex has three neighboring antivortices and vice versa.  
\footnote{Changing the sign of the magnetic field reverses the vorticity for a given sign of 
charge; in this letter we assume that the magnetic field is in the $+z$ direction, {\it i.e.}, $B = B_z > 0$.}

In Fig.~\ref{fig:real space}(c), the effective gap 
$\Delta E$ has been lowered relative to that in Fig.~\ref{fig:real space}(b).
As a result the electron and hole densities are increased 
and the partitioning of the unit charge between the vortex and antivortex core regions is changed.
% Similar to the neutral case, there are intervals of
% $\Delta E$ over which either electrons or 
% holes have an integer filling factor.
By tuning $\Delta E$, we can make the charge near one sublattice approach
one while the charge near the other sublattice approaches zero
so that the honeycomb vortex lattice states continuously evolve to
triangular lattice electron or hole Wigner crystals\footnote{
By hole Wigner crystals, on the electron doping side ($\nu_c>0$),
we refer to the crystal formed by orbitals unoccupied by holes 
in the valence band Landau levels. 
These quasiparticles have the same charge as electrons and 
will form Wigner crystals with the same period. 
See Ref.\cite{macdonald1985broken}.}.
As in the neutral case, in some intervals of $\Delta E$ 
either electrons or holes are at an integer filling factor and excitonic order is lost.
% Compared to Wigner crystals, the charges in vortex lattices fractionalize into 
% two layers and two sublattices.  

% It is intriguing to contrast our vortex lattice state with the Abrikosov 
% vortex lattice state of type-II superconductors. 
% In both systems, introducing a single vortex 
% results in a loss of pair condensation in the vortex core region and
% an increase in pair kinetic energy outside the vortex core.
% In the type II superconductors, the external vector potential contribution to the Cooper pair  
% kinetic momentum approximately cancels the phase variation contribution at large distances resulting in a finite free energy cost. This cost becomes negative beyond 
% a critical magnetic field strength and the ground state ultimately has a finite vortex density that is determined by balancing magnetic and kinetic energies.
% In contrast, the neutral excitons we discuss here do not couple directly 
% to the external vector potential and the ground states of exciton condensates 
% must therefore have zero total vorticity.

\begin{figure}
    \centering
    \includegraphics[width=\linewidth]{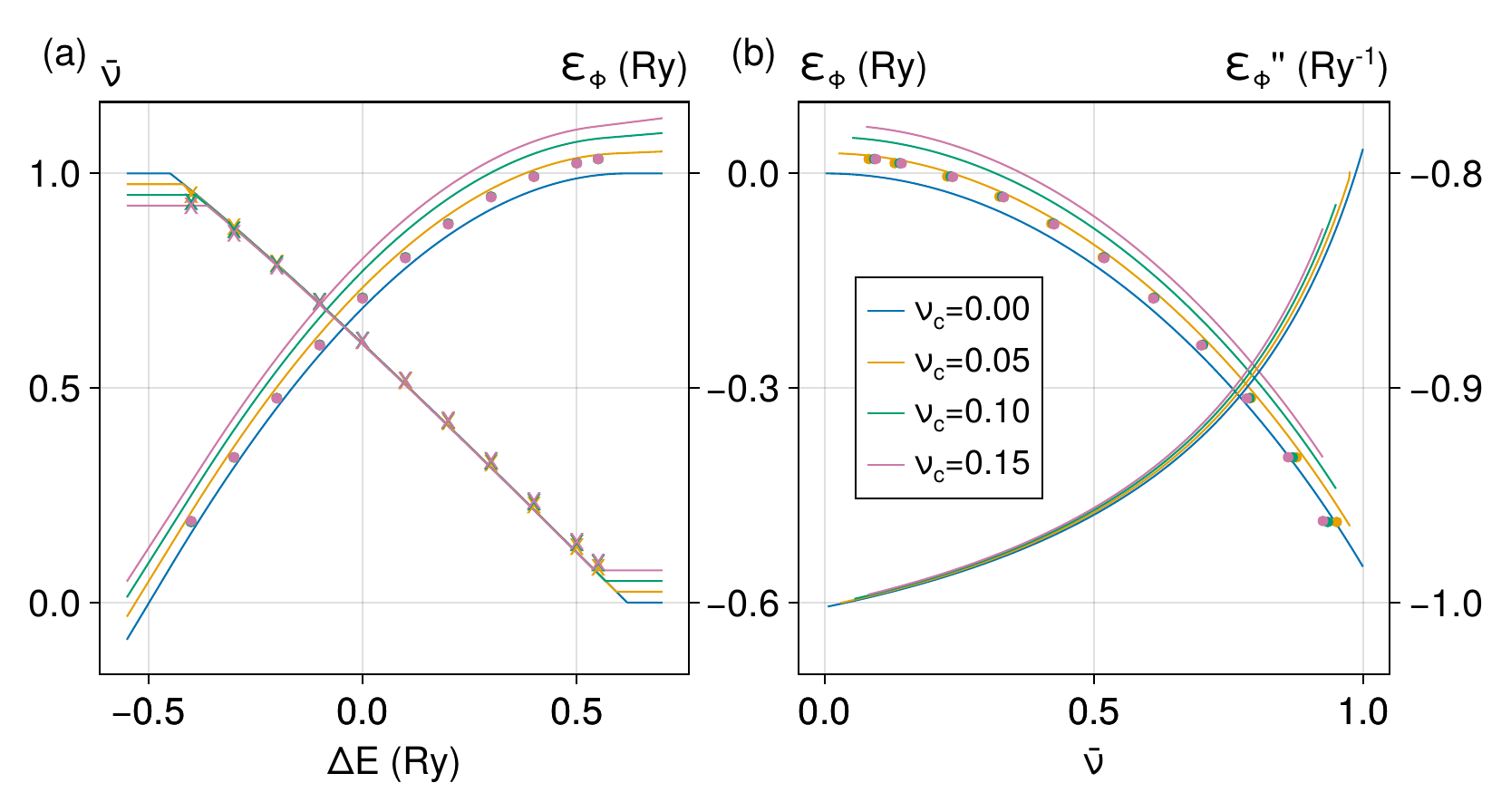}
    \caption{(a)Ground state energy $\mathcal{E}_\Phi$ per $A_\Phi$ unit of area
    and average carrier filling factor
    $\bar{\nu}$ at $B$=0.1$B_0$, $d$=$a_B$ {\it vs.} $\Delta E$. 
    The energies are plotted as circles and the filling factors as crossings.
    The dependence on $\nu_c$, indicated by color, is weak so that markers of different colors often overlap.
    The lines are data for uniform exciton condensates for reference.
    (b)$\mathcal{E}_\Phi$ and its second derivative with respect to $\Delta E$ are plotted 
    {\it vs.} $\bar{\nu}$. The second derivative defines the interaction parameter $U$ in the Bose-Hubbard lattice model approximation to the vortex solutions.  (See the main text.) 
    }
    \label{fig:thermodynanics}
\end{figure}

{\em Quantum Fluctuations and Quantum Melting---}
To account for the role of the order-parameter quantum fluctuations, 
we construct a simplified lattice model that captures the essence of our
microscopic results.  We view the system as consisting of weakly linked condensate regions, 
centered on the triangular lattice sites marked by blue arrows in Fig.~\ref{fig:latticeschematic},
described by a generalized Bose-Hubbard model Hamiltonian
\begin{equation}
\label{eq:latticemodel}
\begin{aligned}
    H = \frac{U}{2} \sum_i  \hat{n}_i(\hat{n}_i-1) - J  \sum_{<ij>} (e^{iA_{ij}} b^\dagger_j b_i + h.c.) ,
\end{aligned}
\end{equation}
where $\left<ij\right>$ are nearest-neighbor sites,
$b_i^{\dagger}$ and $b_i$ are exciton creation and annihilation operators on lattice site $i$ and  
$\hat{n}_i=b_i^{\dagger}b_i$.  
In Eq.~\ref{eq:latticemodel} $A_{ij}$ is an emergent gauge field that we elaborate on below,
$U$ is the on-site exciton-exciton interaction, and $J$ is the Josephson coupling between nearest neighbor condensate regions.

We estimate the strength of $U$ by noticing that $-\Delta E$ acts as a chemical potential 
for excitons so that the inverse short-range interaction strength satisfies 
$(UA_{ex})^{-1} = - \partial^2 \mathcal{E}/\partial \Delta E^2 = -\mathcal{E}^{\prime \prime}$
where $\mathcal{E}$ is the ground state energy per area and $A_{ex}$ is the area over which
the excitons on a given lattice site are spread.
$A_{ex}$ is expected to be smaller than but close to $A_{uc} = |\nu_c|^{-1} A_{\Phi}$, implying that $U \sim  - |\nu_c| / (\mathcal{E}^{\prime\prime} A_{\Phi})$. 
Using the mean-field results plotted in 
Fig.~\ref{fig:thermodynanics}(b) we conclude that 
$ U \sim  |\nu_c| \, {\rm Ry} $ for $B=0.1B_0$ and $d=a_B$.
The Josephson coupling is related to the stiffness of the continuum model \cite{SM}
and is estimated as $J \sim 10^{-2} |\nu_c| {\rm Ry}$ for $B=0.1B_0$ and $d=a_B$.
It follows that $J/U\sim10^{-2}$ is independent of $\nu_c$.

As illustrated in Fig.~\ref{fig:latticeschematic} and Fig.~\ref{fig:real space}, 
the mean-field state has a pattern of phase variation that is induced by the vortices, 
making it different from the normal Bose-Hubbard model.
The emergent gauge field $A_{ij}$ is introduced so that the lattice-model superfluid state 
mimics the continuum HF results, {\it i.e.}, to induce the
differences in exciton consensate phases between nearest-neighbor sites 
induced by the vortices and antivortices.
In the three directions that we plot as gold arrows in Fig.~\ref{fig:latticeschematic},
$A_{ij} = 2\pi/3$, and in the opposite directions, $A_{ij} = -2\pi/3$.
The origin of the gauge field is that the currents around 
the vortices induce effective fluxes - 
$+1$ in the green triangles associated with vortices  
and $-1$ in the orange triangles associated with antivortices, 
as illustrated in Fig.~\ref{fig:latticeschematic}.
The gauge fields simply shift the single-exciton bands
rigidly in momentum space so that the 
excitons condense at $K$ or $K^\prime=-K$ (the two inequivalent corners of the Brillion zone), depending on the sign of $\nu_c$. 
The phase diagram associated with Eq.~\ref{eq:latticemodel} is therefore 
identical to that of the normal Bose-Hubbard \cite{fisher1989boson} model.

The superfluid/insulator phase boundary of the Bose-Hubbard model depends on the 
number of excitons per site $\left<n_i\right>=\nu_{ex}/\nu_c$
and the ratio of Josephson coupling to interaction strength $J/U$.
The vortex-lattice superfluid phase survives
when $J/U$ is larger than $\sim \left<n_i\right>^{-1} \sim
|\nu_c|$ \cite{fisher1989boson}. 
We conclude that the vortex lattice states are stable at small 
$|\nu_c|$ but eventually melt at larger charge densities where 
quantum fluctuations destroy the superfluid order.  
Provided that the broken translation symmetry survives in this 
vortex-fluid state the resulting state could consist of a   
lattice of trions, or more generally of the 
multi-exciton charged complexes anticipated in Ref.~\cite{palacios1996long}. 
We estimate that the critical charge filling factor is 
around $0.1$ for $B=0.1B_0$ and $d=a_B$ and decreases with increasing magnetic field.

\begin{figure}
    \centering
    \includegraphics[width=\linewidth]{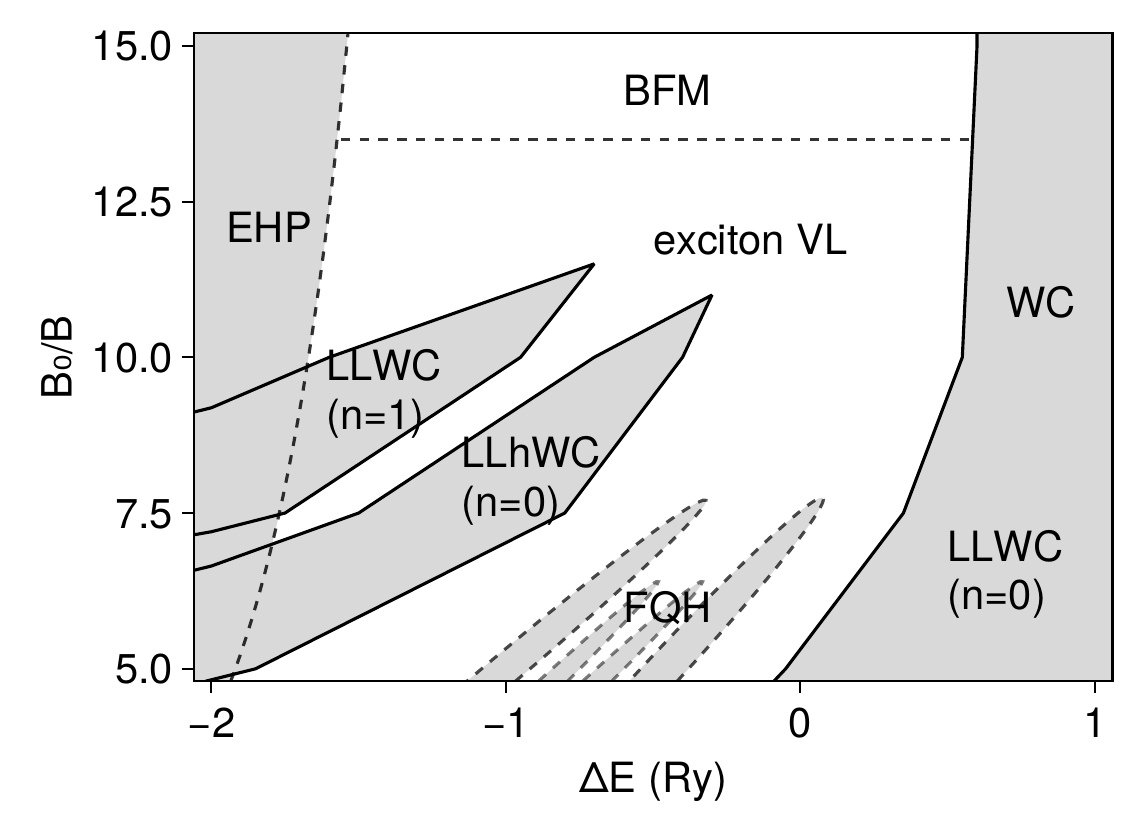}
    \caption{
    Schematic phase diagram of a charged electron-hole fluid at a strong magnetic field $B$ 
    and small positive charge filling factors $\nu_c$.
    The exciton density decreases when the electrically tunable effective band gap $\Delta E$
    increases.  At large positive $\Delta E$, no excitons are present and the ground state is an 
    electron Wigner crystal.  At strong fields, layer-incoherent LLWC and LLhWC 
    states with integer electron or hole filling factors interrupt the VL state and produce magnetic oscillations.
    Note that the Wigner crystal and vortex lattice states share the same lattice constants.  
    The boundaries between these phases have been determined by performing unrestricted  
    HF calculations with $\nu_c=0.1$ and $d=a_B$ - and are plotted as solid lines.
    Phases anticipated in various extreme limits are also shown in this diagram along 
    with estimated phase boundary positions (marked by dashed lines).  
    Fractional incompressible states are anticipated under stronger magnetic fields if the sample is clean. A Mott transition to an electron-hole plasma(EHP) phase is expected at 
    high densities of electrons and holes.
    In weaker fields, charge is not bound to exciton textures, making the crystal easier to melt.
    At both weaker fields and higher $\nu_c$, we expect boson/fermion mixtures (BFM) (in either fluid or crystal states) that contain excitons and charged fermionic exciton complexes to
    compete.  The white regions in this phase diagram have interlayer coherence. 
    }
    \label{fig:phase diagram}
\end{figure}

{\em Discussion---}
In Fig.~\ref{fig:phase diagram}, we propose a schematic phase diagram for 
strong-magnetic-field states of weakly charged electron-hole fluids that goes beyond the   
vortex lattice (VL) and Wigner crystal (WC) states accessible in
mean-field theory by accounting for competing (electron-hole plasmas (EHP), 
Bose-Fermi mixture (BFM), and fractional quantum Hall (FQH) states) expected
to be stable in different extreme limits.  In Fig.~\ref{fig:phase diagram} 
we mark explicitly calculated phase boundaries by solid lines 
and phase boundaries expected due to beyond mean-field physics
by dashed lines placed at crudely estimated positions.
Fig.~\ref{fig:phase diagram} extends the phase diagram in Ref.~\cite{zou2024electrical} 
from neutral to charged systems. 
If we assume that the excess charges are electrons,
electron WC's form at all field strengths for large positive $\Delta E$.
The intermediate $\Delta E$ region is dominated by the exciton-condensate VL states on
which we focus.  These states are counterflow superfluids when vortices are pinned by weak disorder
and should be easy to identify experimentally by large transport drag ratios.
At small $\Delta E$ (larger exciton density), interlayer phase coherence is lost
due to kinetic energy quantization when either $\nu_e$ or $\nu_h$ is an integer.
The vortex lattice then evolves into Landau-level electron or hole Wigner 
crystal (LLWC/LLhWC) states in which the excess charges are accommodated 
by electron or hole Wigner crystals in the partially filled Landau level. 
The Wigner crystals and vortex lattices have the same lattice constants.
Higher-integer-filling (see \cite{SM}) versions of these states 
(not explicitly indicated in Fig.~\ref{fig:phase diagram}) 
are expected at smaller $\Delta E$
until the exciton density exceeds the Mott limit and EHP states emerge.
Fractional-filling analogs of these states are expected in stronger fields in clean samples.  
At weaker fields, mixing of higher Landau levels
is reflected in the spatial patterns of the electron-hole pair amplitude 
and charge density \cite{SM}.
In the weak field limit, vortices do not appear in the ground state.
In this regime we expect that Bose-Fermi mixture (BFM) fluids,
and crystals composed of excitons and electrons, or trions and other 
multi-exciton-electron complexes\cite{palacios1996long} compete.

Our calculations point to the richness of charged 
electron-hole systems in the quantum Hall regime and motivate further experimental study.
The large magnetic field scale $B_0$ for devices realized in TMD 
materials\cite{ ma2021strongly, gu2022dipolar, nguyen2025perfect, zeng2023exciton, qi2025perfect, qi2023thermodynamic, qi2025competition, nguyen2025quantum},
which can be traced to their larger carrier effective masses,
is the main obstacle to the experimental realization of the 
exciton-superfluid vortex lattice state.
The strong field limit of our theory applies equally well to 
graphene electron-electron fluids near total filling factor 1,\cite{li2017excitonic, liu2017quantum, lin2022emergence} where
the required magnetic field scale is much smaller because of the large cyclotron frequency. 
Both vortex lattice and uniform density exciton condensates 
can support counterflow superfluidity.
The vortex lattice state can be distinguished by counterflow-current-driven 
depinning transitions that allow vortices and antivortices to flow and 
provide a dissipation channel. 
Recent developments in scanning tunnel microscopy suggest that 
it might be possibile to measure the spatial pattern of the vortex lattice directly.

{\em Acknowledgements.}---
We thank Kin Fai Mak, Ruishi Qi, Jie Shan, Emanuel Tutuc and Feng Wang for helpful discussions. 
This work is supported by the Office of Naval Research under the Multidisciplinary University Research Initiatives (grant no. N00014-21-1-2377).

\bibliography{vortexlattice.bib}

\clearpage

\section{Supplementary materials}

\setcounter{figure}{0}
\setcounter{equation}{0}
\setcounter{page}{1}
\makeatletter 
\renewcommand{\thefigure}{S\@arabic\c@figure}
\renewcommand{\theequation}{S\@arabic\c@equation}
\makeatother

\subsection{Device geometry and electrostatic potential}

\begin{figure}[b]
    \centering
    \includegraphics[width=\linewidth]{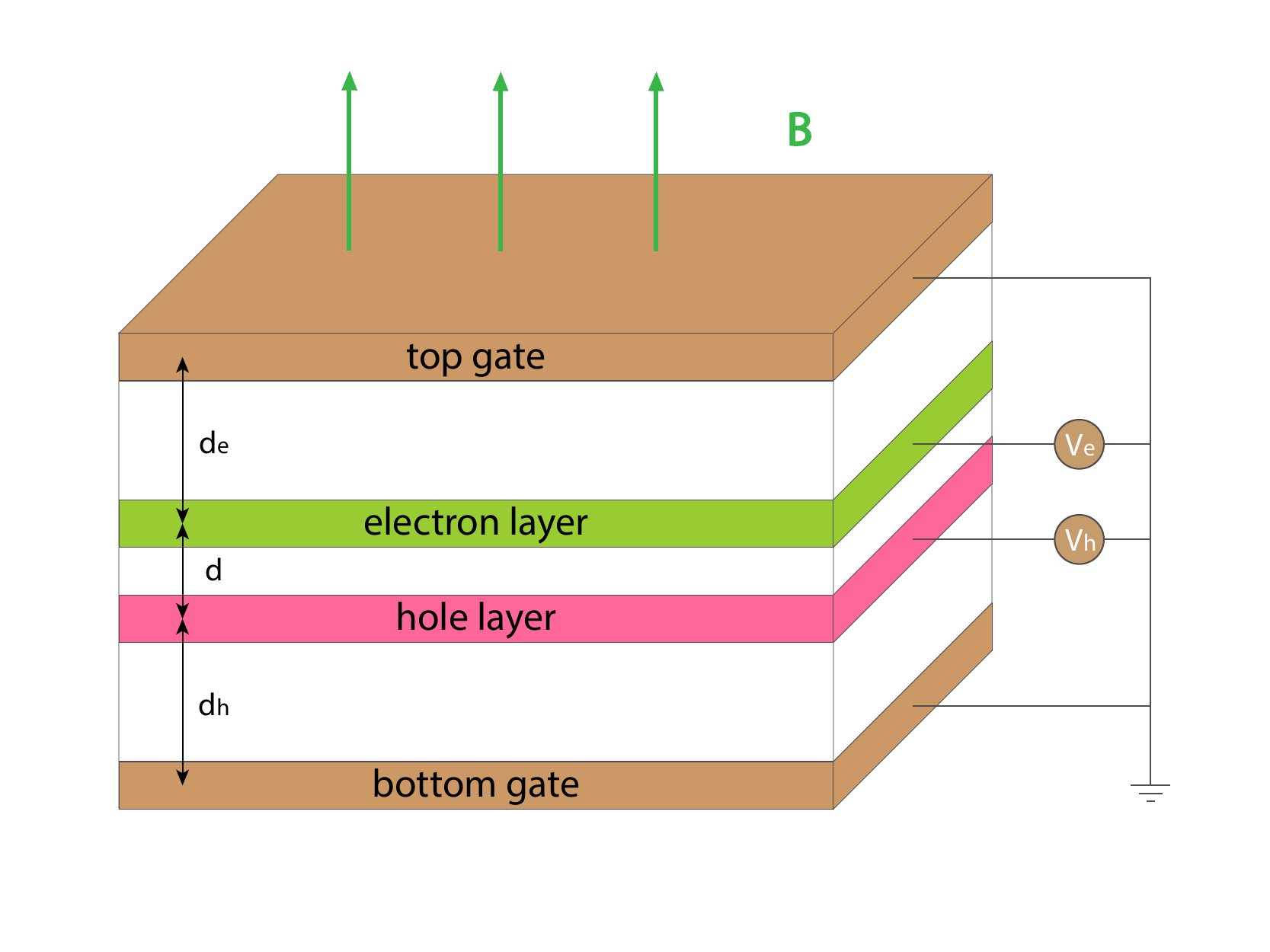}
    \caption{Dual-gated device geometry.}
    \label{fig:device}
\end{figure}

The device geometry analyzed in the main text is shown in Fig.\ref{fig:device}. 
Two semiconducting layers are doped with electrons and holes with 
densities $n$ and $p$ by connecting them to separate reservoirs 
with voltages $V_e$ and $V_h$, respectively.
The two layers are separated by a layer of hexagonal boron nitride (hBN) with
a thickness $d$ large enough that interlayer leakage currents can be neglected.
Because of the opaque hBN barrier
each layer comes to equilibrium with its own reservoir: 
\begin{equation}  \label{eq:equilibrium}
    \begin{aligned}
        -e(V_e - \phi_{e\,}) &=\ \ \epsilon_{c\,}  + \mu_e(n,p), \\
        e(V_h - \phi_h) &= -\epsilon_{v}  + \mu_h(n,p). \\
    \end{aligned}
\end{equation} 
The electric potentials $\phi_{e,h}$ have been explicitly separated in these expressions 
because they are gate geometry dependent, 
and the many-body chemical potentials $\mu_{e,h}$ are measured from the conduction band bottom $\epsilon_c$ and the valence band top $\epsilon_v$ respectively.

Typically the top and bottom gates are separated by large distances 
$d_e$ and $d_h$ and held at the same voltage - which we choose as ground.
For this geometry the electrostatic potential energy
\begin{equation} \label{eq:full electrostatics}
    \begin{aligned}
        -e\,\phi_e(n,p) &= \frac{4\pi e^2 d_e}{\epsilon} \frac{(d_h+d)(n-p)+dp}{d_e+d_h+d}\\
        &\approx \frac{2\pi e^2}{\epsilon}\big[d_g(n-p)+dp\big],\\
        e\,\phi_h(n,p) &= \frac{4\pi e^2 d_h}{\epsilon} \frac{(d_e+d)(p-n)+dn}{d_e+d_h+d}\\
        &\approx \frac{2\pi e^2}{\epsilon}\big[d_g(p-n)+dn\big].
    \end{aligned}
\end{equation}
The simplified versions of these equations assume $d_e=d_h=d_g \gg d$.
Here, $\epsilon$ is the $zz$-element of the hBN dielectric constant.
(We use the simplified formulas in the mean-field calculations explained in the following sections.)
Note that the first term proportional to $d_g$ is fixed by the charge density $n-p$ and
can be neglected in finding the ground state. 
The second termcorresponds to choosing $V(\textbf{q}=0)=0$ for intralayer interactions 
$V(\textbf{q}=0)\sim -2\pi e^2 d/\epsilon$ for the $\textbf{q}=0$ interlayer 
electron-electron interactions. 
These choices are adopted below in Eq.~\ref{eq: hartree}.
Although the term proportional to $d_g$ is irrelevant in assessing the competition between
competing states at a fixed charge density,
it is responsible for the thermodynamic stability of the system, which requires that 
$\partial^2 E / \partial \nu_c ^2 > 0$.
Gate screening effects are important for small momentum $q<1/d_g$.
In our explicit calculations we
assume that $d_g$ is large enough that these can be neglected.

\subsection{Hartree-Fock theory for electron-hole fluids in strong magnetic fields}

We employ the Landau gauge, which is convenient for the description of crystal states,
in our calculations.  The states within a Landau level 
are therefore labeled by the guiding center positions
on the x-axis $X$.  For magnetic fields in the $+\hat{z}$-direction,
the corresponding wavefunctions are 
\begin{equation} \label{eq: wavefunctions in Landau gauge}
    \begin{aligned}
        &\left<\textbf{r}\middle|n\,X\right> = \\
        & \frac{1}{\sqrt{2^n n! L_y}}\left(\frac{1}{\pi l^2} \right)^\frac{1}{4} e^{- \frac{1}{2}(\frac{x-X}{l})^2 + ik_yy} H_n\left(\frac{x-X}{l}\right),
\end{aligned}
\end{equation}
where $n$ is the level index, $l = (\hbar c/eB)^\frac{1}{2}$ is the magnetic length and $k_y = -X/l^2$. The Landau gauge form factors
\begin{equation} \label{eq: Fourier kernel}
    \begin{aligned}
        \left< n_1\,X_1 \middle| e^{i\textbf{q} \cdot \textbf{r}} \middle| n_2\, X_2 \right> = \ \ & \\
        F_{n_1,n_2}&(\textbf{q}) e^{iq_x\frac{X_1+X_2}{2}} \delta_{X_1,X_2-q_yl^2},
    \end{aligned}
\end{equation}
where 
\begin{equation} \label{eq: form factor}
\begin{aligned}
    &F_{n_1,n_2}(q_x,q_y) = F_{n_1,n_2}(q,\theta_\textbf{q})\\
    =&\left\{
    \begin{aligned}
        \sqrt\frac{n_1!}{n_2!} \left(\frac{iq_x-q_y}{\sqrt{2}}l\right)^{n_2-n_1} \times&\\
        e^{-\frac{q^2l^2}{4}} L_{n_1}^{(n_2-n_1)}&\left(\frac{q^2l^2}{2}\right),n_1\le n_2\\
        \sqrt\frac{n_2!}{n_1!} \left(\frac{iq_x+q_y}{\sqrt{2}}l\right)^{n_1-n_2} \times&\\
        e^{-\frac{q^2l^2}{4}} L_{n_2}^{(n_1-n_2)}&\left(\frac{q^2l^2}{2}\right),n_1\ge n_2\\
    \end{aligned}
    \right.\\
    =& \sqrt\frac{n_<!}{n_>!} \left(\frac{iql}{\sqrt{2}}\right)^{n_>-n_<} e^{i\theta_\textbf{q}(n_2-n_1)}\ \times \\
    &\ \ \ \ \ \ \ \ 
    e^{-\frac{q^2l^2}{4}} L_{n_<}^{(n_>-n_<)}\left(\frac{q^2l^2}{2}\right),
\end{aligned} 
\end{equation}
$(q,\theta_\textbf{q})$ are the polar coordinates of $\textbf{q}$, 
% Specifically we choose $\theta_\textbf{q}$ to be zero along the
% $+x$ axis and to increase in the counterclockwise direction.
$n_>$ and $n_<$ are the larger and smaller integers 
among $(n_1,n_2)$, and $L_n^{(\alpha)}(x)$ is a generalized Laguerre polynomial. 
Note that $F_{n_1,n_2}(q_x=0,q_y=0) = \delta_{n_1,n_2}$.
The momentum-space density operator is therefore 
\begin{equation} \label{eq: n(q)}
\begin{aligned}
        n_{b b^\prime}(\textbf{q})=&\int{d\textbf{r}}\,e^{-i\textbf{q}\cdot\textbf{r}} \psi^\dagger_{b^\prime}(\mathbf{r}) \psi_{b}(\mathbf{r})\\
        =&\sum_{n n^\prime}\sum_{X X^\prime} \left< n^\prime\,X^\prime \middle| e^{-i\textbf{q}\cdot\textbf{r}} \middle| n\,X \right> 
        c^\dagger_{b^\prime, n^\prime, X^\prime} c_{b, n, X}\\
        =& N_\Phi \sum_{n n^\prime}\rho_{\substack{n n^\prime\\b b^\prime}}(\textbf{q})F_{n^\prime,n}(-\textbf{q}).
\end{aligned}
\end{equation}
In Eq.~\ref{eq: n(q)}, $b=c/v$ is the band/layer index, 
$N_\Phi=\Phi/\Phi_0$ is the number of states in each Landau level,
and we have defined the Landau-level-resolved density operator $\rho{\substack{n n^\prime\\ b b^\prime}}(\textbf{q})$ as 
\begin{equation} \label{eq: density matrix (q)}
    \rho_{\substack{n n^\prime\\ b b^\prime}}(\textbf{q}) = \frac{1}{N_\Phi} \sum_X e^{-iq_x (X + \frac{q_y}{2} l^2)} c^\dagger_{b^\prime,n^\prime,X+q_yl^2} c_{b,n,X}.
\end{equation}
The prefactor $1/N_\Phi$ makes $\rho(\textbf{q})$ an intensive quantity. 
Importantly Eq.~\ref{eq: density matrix (q)} can be inverted: 
\begin{equation} \label{eq: density matrix b n X}
    c^\dagger_{b^\prime,n^\prime,X^\prime} c_{b,n,X} = \sum_\textbf{q} \delta_{X^\prime, X+q_yl^2} e^{iq_x \frac{X+X^\prime}{2}} \rho_{\substack{n n^\prime\\ b b^\prime}}(\textbf{q}).
\end{equation}
This remarkable formula implies that in the projected Hilbert space of any
Landau level the position dependence of the charge density (the diagonal elements of the density matrix in a coordinate representation) uniquely determines the full density-matrix and therefore
the Fock exchange potential.  Below we use this property, which can be traced \cite{macdonald1988density} to the analytic Hilbert-space of a single Landau-level, 
to simplify the Hartree-Fock calculations.

% We derive the Hartree-Fock approximation for a Landau level system by calculating the 
% equation of motion of the electron annihilation operator.  
% The commutator between annihilation operators and density operators
% \begin{equation} \label{eq: [rho(q), c]}
% \begin{aligned}
%     \left[ \rho_{\substack{n n^\prime\\ b b^\prime}}(\textbf{q}) \ ,\ c_{c,m,X} \right] = \ \ \ \ \ \ \ \ \ \ \ &  \\
%     -\frac{1}{N_\Phi} e^{-iq_x (X - \frac{1}{2}q_y l^2)} & c_{b,n,X-q_yl^2} \delta_{b^\prime c} \delta_{n^\prime m}.
% \end{aligned}
% \end{equation}
We define $V_{bb^\prime}(\textbf{q}) = 2\pi e^2 \exp(-qd(b,b^\prime))/\epsilon q$ 
as the two-dimensional Fourier transform of the Coulomb energy between electrons 
in layers $b$ and $b^\prime$, separated by distance $d(b,b^\prime)$. 
With this notation the interaction Hamiltonian
\begin{equation} \label{eq: interaction}
\begin{aligned}
    V = \frac{1}{2A}& \sum_{b\,b^\prime} \sum_\textbf{q} \sum_{\substack{n_1,n_2,n_3,n_4\\X_1,X_2,X_3,X_4}} V_{bb^\prime}(\textbf{q}) \\
    & \left< n_1 X_1 \middle| e^{-i\textbf{q}\cdot\textbf{r}} \middle| n_4 X_4 \right> \left< n_2 X_2 \middle| e^{i\textbf{q}\cdot\textbf{r}} \middle| n_3 X_3 \right>\\
    &c^\dagger_{b,n_1,X_1} c^\dagger_{b^\prime,n_2,X_2} c_{b^\prime,n_3,X_3} c_{b,n_4,X_4}.\\
\end{aligned}
\end{equation}
Note that the layer indices are conserved at each vertex.
Using Eq.~\ref{eq: density matrix b n X} we find that the 
Hartree-Fock (HF) Hamiltonian can be expressed in terms of density operators:
\begin{equation} \label{eq: hamiltonian HF}
\begin{aligned}
    H=& N_\Phi\sum_{nb} \epsilon_{n,b}\,\rho_{\substack{n n\\b\,b}}(\textbf{q}=0)  + V\\
    =& N_\Phi\sum_{nb} \epsilon_{n,b}\,\rho_{\substack{n n\\ b\,b}}(\textbf{q}=0)\\
        &\ \ \ \ \ \ \ \ \ 
    +N_\Phi \sum_{\substack{n n^\prime \, b b^\prime}} \sum_\textbf{q} U_{\substack{n^\prime n\\b^\prime b}}(\textbf{q})\ \rho_{\substack{n n^\prime\\b b^\prime}}(\textbf{q})\,,
\end{aligned}
\end{equation}
where the coupling function $U$ contains Hartree($H$) and Fock/exchange($X$) contributions,
$U=U^H - U^X$, defined below
In these equations the density matrix is understood to be regularized by 
subtracting from it the full valence band density-matirx,
$\delta_{bv}\delta_{bb^\prime}\delta_{nn^\prime}\delta_{\textbf{q}0}$,
since the single-particle energies implicitly include the self-energies 
they produce.  

Both $U^H$ and $U^{X}$ are linear in the self-consistent 
density matrix $\big< \rho{\substack{n n^\prime\\ b b^\prime}}(\textbf{q}) \big>$. 
The Hartree contribution to $U$ is diagonal in layer
and involves only layer-diagonal density-matrix elements: 
\begin{equation} \label{eq: UH}
\begin{aligned}
    &U^H {\substack{n^\prime n\\b^\prime b}}(\textbf{q}) 
    = U^H {\substack{n^\prime n\\b\ b}}(\textbf{q}) \, \delta_{b,b^\prime}\\
    =& \frac{N_\Phi}{A} \delta_{b,b^\prime}\sum_{\substack{mm^\prime \, c}} V_{bc}(\textbf{q}) F_{ n^\prime,n}(-\textbf{q}) F_{m^\prime,m}(\textbf{q}) \\
    &\ \ \ \ \ \ \ \ \ \ \ \ \ \ \ \ \ \ \ \ \ \ \ \ \ \ \ \ \ \ \ \ \ \ \ \ \ \ \ \times\Big< \rho_{\substack{m m^\prime\\c\,c}}(-\textbf{q})\Big> \\
    =& W_0 \,\delta_{b,b^\prime} \sum_{mm^\prime} \Bigg( H_{nn^\prime;mm^\prime}(\textbf{q}) \Big< \rho_{\substack{m m^\prime\\b\,b}}(-\textbf{q})\Big> \\
    &\ \ \ \ \ \ \ \ \ \ \ \ \ \ \ \ \ \ \ \ 
    +H^*_{nn^\prime;mm^\prime}(\textbf{q}) \Big< \rho_{\substack{m m^\prime\\ \overline{b}\, \overline{b} }}(-\textbf{q})\Big> \Bigg),
\end{aligned}
\end{equation}
where $W_0=e^2/\epsilon l$ is the Coulomb energy scale, $\overline{b}$ is the
layer opposite to $b$, and the coefficients are given by
\begin{equation} \label{eq: hartree}
\begin{aligned}
        &H_{nn^\prime;mm^\prime}(\textbf{q}\neq0) = \frac{F_{n^\prime, n}(-\textbf{q}) F_{m^\prime, m}(\textbf{q})}{ql},\\
        &H^*_{nn^\prime;mm^\prime}(\textbf{q}\neq0) = \frac{F_{n^\prime, n}(-\textbf{q}) F_{m^\prime, m}(\textbf{q})}{ql} e^{-qd},\\
        &H_{nn^\prime;mm^\prime}(0) = 0\,,\\
        &H^*_{nn^\prime;mm^\prime}(0) = \lim_{q\rightarrow0} H^*_{nn^\prime;mm^\prime}(\textbf{q}) - H_{nn^\prime;mm^\prime}(\textbf{q})\\
        &\ \ \ \ \ \ \ \ \ \ \ \ \ \ \ \ = \delta_{nn^\prime} \delta_{mm^\prime}\, \frac{-d}{l}.
\end{aligned}
\end{equation}
Here the asterisk distinquishes interlayer from intralayer interaction contributions.
As mentioned above, our treatment of Coulomb interaction energy at $\textbf{q}=0$
assumes that $d_g \gg d$ (see Eq.~\ref{eq:full electrostatics}),
and as explained previously leads to the convention that the same layer 
interaction $V_{bb}(0)=0$.

The exchange contribution $U^X$ has contributions that are off-diagonal in layer that 
are proportional to layer-off-diagonal density matricx components: 
\begin{equation} \label{eq: UX}
\begin{aligned}
    &U^X {\substack{n^\prime n\\b^\prime b}}(\textbf{q}) \\
    =& \sum_{\substack{mm^\prime}}\int \frac{d\,\textbf{k}}{(2\pi)^2}  V_{bb^\prime}(\textbf{k}) F_{ n^\prime,m}(-\textbf{k}) F_{m^\prime,n}(\textbf{k}) \, e^{i\textbf{q}\times\textbf{k}l^2}\\
    &\ \ \ \ \ \ \ \ \ \ \ \ \ \ \ \ \ \ \ \ \ \ \ \ \ \ \ \ \ \ \ \ \ \ \ \ \ \ \ \ \times \Big< \rho_{\substack{m m^\prime\\b^\prime\,b}}(-\textbf{q})\Big> \\
    =& \left\{
    \begin{aligned}
        & W_0 \sum_{\substack{mm^\prime}} X_{nn^\prime;mm^\prime}(\textbf{q}) \Big< \rho_{\substack{m m^\prime\\ b^\prime\, b }}(-\textbf{q})\Big>, \ b=b^\prime \\
         & W_0 \sum_{\substack{mm^\prime}} X^*_{nn^\prime;mm^\prime}(\textbf{q}) \Big< \rho_{\substack{m m^\prime\\ b^\prime\, b }}(-\textbf{q})\Big>, \ b=\overline{b^\prime}\, . \\
    \end{aligned}
    \right.
\end{aligned}
\end{equation}
Defining 
\begin{equation*}
\begin{aligned}
    &s_1 = m^\prime -n,\ n_{1>} = \text{max}\{m^\prime,n\},\ n_{1<} = \text{min}\{m^\prime,n\},\\
    &s_2 = n^\prime -m,\ n_{2>} = \text{max}\{n^\prime,m\},\ n_{2<} = \text{min}\{n^\prime,m\},\\
\end{aligned}
\end{equation*}
the coefficients $X$ and $X^*$ are given by 
\begin{equation} 
\begin{aligned}
    &X_{nn ^\prime;mm^\prime}(q,\theta_\textbf{q})= i^{|s_1| - |s_2|} e^{-i(s_1+s_2)\theta_\textbf{q}}  \\
    &\ \ \ \ \times \sqrt\frac{n_{1<}!\,n_{2<}!}{n_{1>}!\,n_{2>}!} \int_0^{\infty} e^{-\frac{x^2}{2}} \left(\frac{x^2}{2}\right)^ {\frac{1}{2}(|s_1|+|s_2|)}\\
    &\ \ \ \ \ \ \ \ \ L_{n_{1<}}^{(|s_1|)}\left(\frac{x^2}{2}\right) L_{n_{2<}}^{(|s_2|)}\left(\frac{x^2}{2}\right) J_{s_1+s_2}\left(qlx\right) dx,\\
\end{aligned}
\end{equation}
\begin{equation}
\begin{aligned}
    &X^*_{nn ^\prime;mm^\prime}(q,\theta_\textbf{q})= i^{|s_1| - |s_2|} e^{-i(s_1+s_2)\theta_\textbf{q}}  \\
    &\ \ \ \ \times \sqrt\frac{n_{1<}!\,n_{2<}!}{n_{1>}!\,n_{2>}!} \int_0^{\infty} e^{-\frac{x^2}{2}-x\frac{d}{l}} \left(\frac{x^2}{2}\right)^ {\frac{1}{2}(|s_1|+|s_2|)}\\
    &\ \ \ \ \ \ \ \ \ L_{n_{1<}}^{(|s_1|)}\left(\frac{x^2}{2}\right) L_{n_{2<}}^{(|s_2|)}\left(\frac{x^2}{2}\right) J_{s_1+s_2}\left(qlx\right) dx,\\
\end{aligned}
\end{equation}
where the $J_n(x)$ are Bessel functions of the first kind.
Two properties
\begin{equation}
\begin{aligned}
    & X^{(*)}_{m^\prime m;n^\prime n}(\textbf{q}) = \big[ X^{(*)}_{nn^\prime;mm^\prime}(\textbf{q}) \big]^* \\
    & X^{(*)}_{mm^\prime;nn^\prime}(\textbf{q}) = \, X^{(*)}_{nn^\prime;mm^\prime}(-\textbf{q})\  \\
\end{aligned}
\end{equation}
are useful in reducing the number of integrals that have to be evaluated numerically;
the first ensures that the Hamiltonian is Hermitian, 
and the second originates from invariance of the Coulomb interaction under interchange of 
particle indices.  The energy per flux quantum 
\begin{equation}
\begin{aligned}
    \mathcal{E}_\Phi = \frac{\left<H\right>}{N_\Phi}
    =& \sum_{nb} \epsilon_{n,b}\Big< \rho_{\substack{n n\\ b\,b}}(\textbf{q}=0)\Big>\\
        &\ \ \ \ 
    +{ \frac{1}{2} } \sum_{\substack{n n^\prime \, b b^\prime}} \sum_\textbf{q} U_{\substack{n^\prime n\\b^\prime b}}(\textbf{q})\Big< \rho_{\substack{n n^\prime\\b b^\prime}}(\textbf{q})\Big>,
\end{aligned}
\end{equation}
and energy per area is $\mathcal{E} = \mathcal{E}_\Phi/A_{\Phi}$.

\subsection{Equation-of-motion Hartree-Fock solution}

We define the momentum space Matsubara Green's functions as 
\begin{equation} \label{eq: green's function}
\begin{aligned}
    G{\substack{n n^\prime\\ b b^\prime}}(\textbf{q}; \tau) = \frac{1}{N_\Phi} & \sum_X e^{-iq_x (X + \frac{1}{2}q_y l^2)}\ \times \\
    & \ \left<\text{T}\,\Big[c^\dagger_{b^\prime,n^\prime,X+q_yl^2}(0) c_{b,n,X}(\tau)\Big]\right>
\end{aligned}
\end{equation}
has the same form as the density operator in Eq.~\ref{eq: density matrix (q)}.
Here the imaginary-time creation and annihilation operators $c^{(\dagger)}_{b,n,X}(\tau) = e^{\tau (H-\mu N)} c^{(\dagger)}_{b,n,X} e^{-\tau (H-\mu N)}$, where $\mu$ is the chemical potential and
$\text{T}$ indicated time-ordering.
The commutator of an annihilation operator with the density operator
\begin{equation}
\begin{aligned}
    \Big[ \rho_{\substack{n n^\prime\\ b b^\prime}}(\textbf{q}) \ ,\ c_{c,m,X} \Big] = \ \ \ \ \ \ \ \ \ \ \ &  \\
    -\frac{1}{N_\Phi} e^{-iq_x (X - \frac{1}{2}q_y l^2)} & c_{b,n,X-q_yl^2} \delta_{b^\prime c} \delta_{n^\prime m}.
\end{aligned}
\end{equation}
It follows that the commutator with the Hamiltonian in Eq.~\ref{eq: hamiltonian HF}
\begin{equation}
\begin{aligned}
    \left[H-\mu N \ ,\ c_{c,m,X} \right] = &-  \sum_{nb} (\epsilon_{n,b}-\mu)c_{c,m,X} \\
    -\sum_{n  b } \sum_\textbf{k} U_{\substack{m n\\c\,b}}(\textbf{k}) & e^{-ik_x (X - \frac{1}{2}k_y l^2)}  c_{b,n,X-k_yl^2}.
\end{aligned}
\end{equation}
and that 
\begin{equation} \label{eq: dG/dtau}
\begin{aligned}
    &\hbar \frac{d}{d\tau} G{\substack{n n^\prime\\ b b^\prime}}(\textbf{q}; \tau)\\
    =& \frac{1}{N_\Phi} \sum_X e^{-iq_x (X + \frac{1}{2}q_y l^2)}\ \times \\
    &\ \ \Bigg[ - \hbar\delta(\tau)\left\{ c^\dagger_{b^\prime,n^\prime,X+q_yl^2}\ ,\  c_{b,n,X} \right\} \\
    &\ \ \ \  \left< \text{T}\,c^\dagger_{b^\prime,n^\prime,X+q_yl^2}(0) \left[ H-\mu N, c_{b,n,X} \right](\tau) \right> \Bigg]\\
    =& - \hbar\delta(\tau) \delta_{bb^\prime} \delta_{nn^\prime} \delta_{\textbf{q}0} - (\epsilon_{n,b}-\mu) G{\substack{n n^\prime\\ b b^\prime}}(\textbf{q}; \tau) \\
    &-\sum_{mc,\textbf{k}} U_{\substack{m n\\c\,b}}(\textbf{k}) e^{i\frac{1}{2} \textbf{k}\times\textbf{q} l^2} G{\substack{m n^\prime\\ c b^\prime}}(\textbf{q}+\textbf{k}; \tau),\\
\end{aligned}
\end{equation}
After Fourier transforming to Matsubara frequencies, we obtain
\begin{equation} \label{eq: eom green's function}
\begin{aligned}
    (i\omega_n &+ \frac{\mu}{\hbar}) G{\substack{n n^\prime\\ b b^\prime}}(\textbf{q}; i\omega_n) \\
    -& \sum_{mc,\textbf{q}^\prime} A{\substack{n m\\ b\,c}}(\textbf{q}, \textbf{q}^\prime) G{\substack{m n^\prime\\ c\,b^\prime}}(\textbf{q}^\prime; i\omega_n) = \delta_{bb^\prime} \delta_{nn^\prime} \delta_{\textbf{q}0},
\end{aligned}
\end{equation}
where
\begin{equation} \label{eq: A matrix}
\begin{aligned}
    \hbar A&{\substack{n m\\ b\,c}}(\textbf{q}, \textbf{q}^\prime) = \\
    &\epsilon_{n,b} \delta_{nm} \delta_{bc} \delta_{\textbf{q} \textbf{q}^\prime } + U{\substack{n m\\ b\,c}}(\textbf{q}^\prime - \textbf{q}) e^{i\frac{1}{2}\textbf{q}^\prime \times \textbf{q} l^2}
\end{aligned}
\end{equation}
is a Hermitian matrix which we diagonalize numerically:
\begin{equation}
    \sum_{mc,\textbf{q}^\prime} A{\substack{n m\\ b\,c}}(\textbf{q}, \textbf{q}^\prime) V^{(j)}_{cm}(\textbf{q}^\prime) = \omega^{(j)} V^{(j)}_{bn}(\textbf{q}). 
\end{equation}
% These eigenvectors are normalized in the meaning of
% \begin{equation}
%     \sum_{nb,\textbf{q}} \big[ V^{(j_1)}_{bn}(\textbf{q}) \big]^* \  V^{(j_2)}_{bn}(\textbf{q}) = \delta^{j_1 j_2}.
% \end{equation}
It follows that the Green's function is 
\begin{equation}
    G{\substack{n n^\prime\\ b b^\prime}}(\textbf{q}; i\omega_n) = \sum_j \frac{ V^{(j)}_{bn}(\textbf{q}) \big[ V^{(j)}_{b^\prime n^\prime}(0)  \big]^*  }{i\omega_n + \mu/\hbar - \omega^{(j)}}.
\end{equation}
and that the self-consistent field equation for the density-matirx is  
\begin{equation}
\begin{aligned} \label{eq:Matsubara sum}
    \Big< \rho_{\substack{n n^\prime\\ b b^\prime}}(\textbf{q}) \Big> =& \lim_{\tau\rightarrow 0^-} G{\substack{n n^\prime\\ b b^\prime}}(\textbf{q}; \tau) - \delta_{bv}\delta_{bb^\prime}\delta_{nn^\prime}\delta_{\textbf{q}0}\\
    =& \sum_j n_F( \hbar\omega^{(j)} - \mu) V^{(j)}_{bn}(\textbf{q}) \big[ V^{(j)}_{b^\prime n^\prime}(0)  \big]^* \\
    & -\delta_{bv}\delta_{bb^\prime}\delta_{nn^\prime}\delta_{\textbf{q}0},
\end{aligned}
\end{equation}
where $n_F$ is the Fermi-Dirac distribution function, and the chemical potential $\mu$ is determined by the filing factor
\begin{equation}
\begin{aligned}
    \nu_c = \sum_{bn} \Big< \rho_{\substack{n n\\ b b}}(0) \Big>.
\end{aligned}
\end{equation}

\subsection{Broken Translational Symmetry}

The previous sections explain the basic equations we use to solve unrestricted
Hartree-Fock equations for an electron-hole fluid in a Landau level basis.  
The formulation of Hartree-Fock theory is unusual
and takes advantage of the analyticity of the wavefunctions within each Landau level.  
Because we express the Hartree-Fock equations in terms of 
charge-density Fourier components $\big< \rho{\substack{n n^\prime\\ b b^\prime}}(\textbf{q}) \big>$, we can find broken translational-symmetry solutions with any lattice
periodicity by allowing these to be non-zero only on the corresponding
reciprocal lattice of momenta $\textbf{q}$.  Since the Hamiltonian is translationally
invariant, non-zero $\big< \rho{\substack{n n^\prime\\ b b^\prime}}(\textbf{q}) \big>$ at any
non-zero $\textbf{q}$ signals a broken symmetry.  
Layer-diagonal density-matrix elements are related to charge density fluctuations
whereas layer-off-diagonal elements define the 
real-space pattern of interlayer excitonic coherence.

If we allow $\big< \rho{\substack{n n^\prime\\ b b^\prime}}(\textbf{q}) \big>$ 
to be non-zero only at $\textbf{q}=0$, 
both the charge and the exciton order parameter are uniform in real space.
The vortex lattice solutions we find at small finite $\nu_c$ (small electron-hole imbalance) 
restrict $\textbf{q}$ to a reciprocal lattice corresponding to one charge per unit cell.    
We also find uniform solutions by restricting momenta to $\textbf{q}=0$,
but these competitors to the vortex lattice states 
turn out to always have higher energy, as reported in Fig.(3) in the main text.
Because the charge densities and excitonic order parameters are smooth functions of 
postion we can cut-off the number of shells of the reciprocal lattice vectors.
Our calculations do not assume that vortices and antivortices are present or that 
they have any particular spatial pattern when present, only that the solutions that we 
find are period.  We can find solutions for any lattice type, but we find that
the honeycomb lattice solutions discussed in the main text are lowest in energy.

The ground state solutions discussed in the main text 
have zero total counterflow supercurrent.
We have also found solutions with finite supercurrent by simultaneously 
shifting the allowed momenta by $\textbf{Q}$ relative to reciprocal lattice vectors 
for $\rho_{cv}$ and by $-\textbf{Q}$ for $\rho_{vc}$.
This transformation finds excitons with lattice momentum $\textbf{Q}$. 
We have carried out calculations of this type for both uniform and vortex lattice ground states
and used them to determine the exciton superfluid density discussed
in the next section.

To solve the Hartree-Fock equations, we use  
a trial density matrix to construct the Greens function, and diagonalize it to obtain  
an updated density matrix.  This procedure is repeated until
convergence is reached.  Once the density matrix is converged, we calculate the corresponding
energy and, by inverting the Fourier transform in Eq.~\ref{eq: n(q)}, real space observables.  
Each iteration contains the following steps:
(a) Calculate the mean field $U$ using Eq.~\ref{eq: UH} and ~\ref{eq: UX} and
a weighted average of the input and output density matrices of the previous iteration
(b) Find the $A$ matrix using Eq.~\ref{eq: A matrix};
(c) Diagonalize the $A$ matrix and find its eigenvalues $\omega^j$ and eigenvectors $V^j$;
(d) Calculate the new density matrix $\rho_\text{new}$ with $\omega$ and $V$ using Eq.~\ref{eq:Matsubara sum} with $\mu$ determined by $\nu_c$.

\begin{figure}
    \centering
    \includegraphics[width=\linewidth]{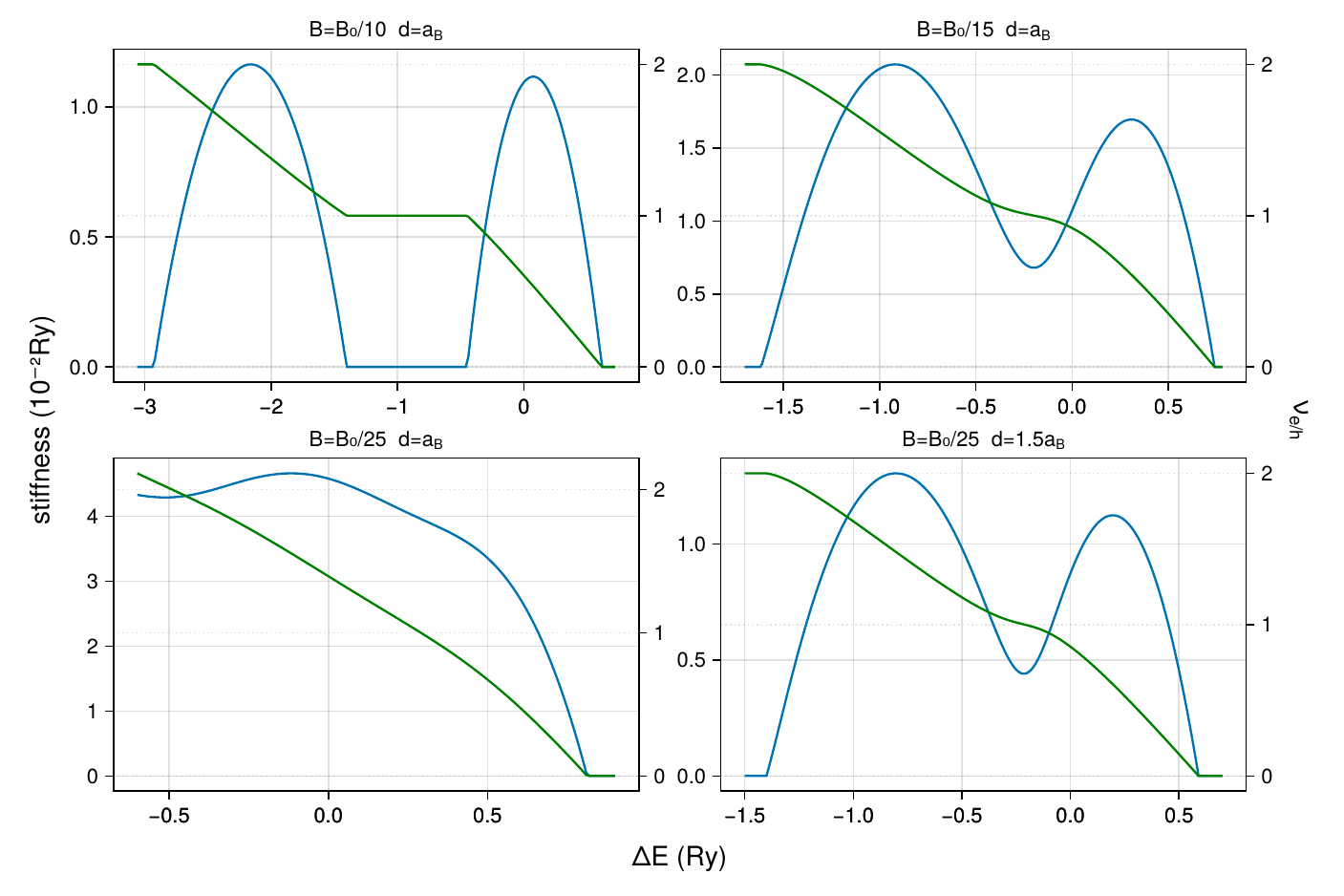}
    \caption{The exciton condensate stiffness (blue) and carrier filling factor (green) at neutrality {\it vs.} band gap $\Delta E$.}
    \label{fig:stiffness}
\end{figure}

\subsection{Exciton superfluid density and the Josephson junction model hopping parameter}
For the stiffness calculation we shift the interlayer coherence momenta by $\pm \textbf{Q}$ 
as explained in the previous equations.
The Hartree-Fock equation solutions can be solved for any value of $\pm \textbf{Q}$ in the 
Brillouin-zone of the vortex lattice state.
The main text focuses on the lowest energy solutions with
coherence momentum $\pm \textbf{Q}=0$, and the expansion of the energy at 
small momentum is istropic to quadratic order for triangular vortex lattices.
The exciton stiffness $s$ of the electron-hole condensate may be   
defined as the coefficient of the quadratic term in the 
small $Q$ expansion of the condensate energy per area: 
\begin{equation}\label{eq:stiffness}
    \mathcal{E}(Q) = \mathcal{E}(0) + s Q^2 + \cdots.
\end{equation}
Fig.~\ref{fig:stiffness} plots the stiffness of neutral 
exciton condensates as blue lines, along with the exciton filling factor,
as green lines.
The stiffness is proportional to the superfluid density $\nu_{ex}$.
In strong fields, it reaches a maximum at half-filling of each Landau level,
{\it i.e.} when $\nu_{ex}=0.5$.

The tight-binding kinetic part of the effective Bose-Hubbard model
(eq.~\ref{eq:latticemodel} in the main text) is 
\begin{equation}
    H = - J \sum_{<ij>} (e^{iA_{ij}} b^\dagger_j b_i + h.c. )
\end{equation}
gives the exciton band dispersion
\begin{equation}
    E_Q = - 2J \sum_{i=1}^3 \cos{(\textbf{Q}\cdot\textbf{a}_i + A(\textbf{a}_i))},
\end{equation}
where $\textbf{a}_{1,2,3}$ are three 
lattice vectors from the first shell with 120-degree relative 
angles and $A(\textbf{a}_i)$ is the hopping phase gained from the gauge field.
In neutral condensates, $A(\textbf{a}_i)=0$ and the band reaches its minimium 
at $\textbf{Q}=0$:
\begin{equation}
    E_Q = E_0 + \frac{\sqrt{3}JA_\Phi}{|\nu_c|} Q^2 + \cdots.
\end{equation}
Therefore, if the excitons are Bose condensed, the total energy per area
\begin{equation} \label{eq:hopping energy}
    \mathcal{E}(Q) = E_Q \frac{\nu_{ex}}{A_\Phi} =  \mathcal{E}(0) + \frac{\sqrt{3}J\nu_{ex}}{|\nu_c|} Q^2 + \cdots.
\end{equation}
Comparing eq.~\ref{eq:stiffness} and ~\ref{eq:hopping energy}, we conclude that 
we should choose 
\begin{equation}
    J = \frac{|\nu_c|}{\sqrt{3}\nu_{ex}}s
\end{equation}
in the lattice model.

\subsection{Extended Results: Higher Landau-level solutions}

\begin{figure*}
    \centering
    \includegraphics[width=\linewidth]{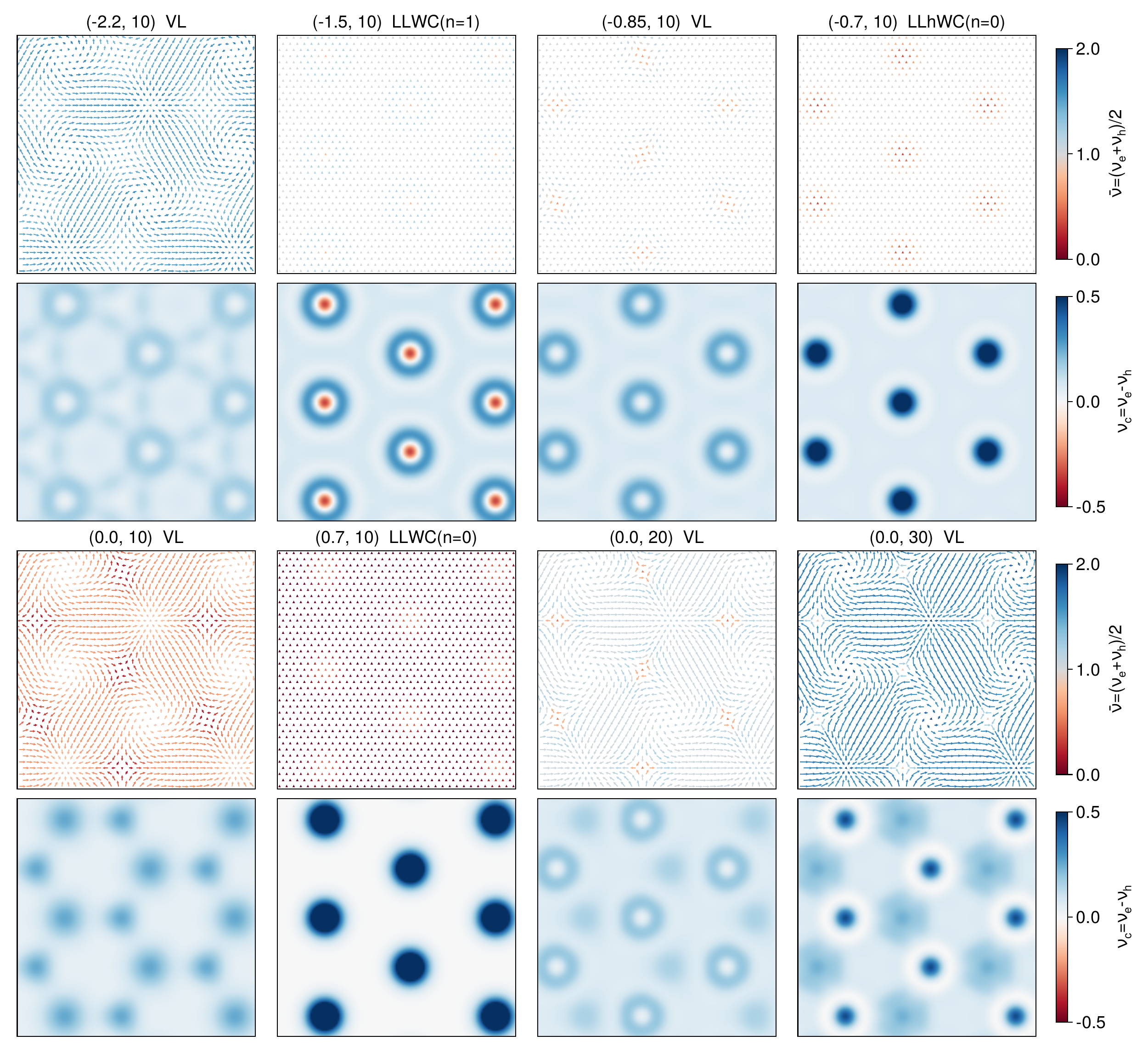}
    \caption{Hartree-Fock calculation result summaries at eight additional 
    ($\Delta E/Ry, B_0/B$) values at $d=a_B$.  For each value, the upper panel illustrates the 
    pairing amplitude and phase and the bottom panel charge density.
    The layout of these figures is similar to that of Fig~\ref{fig:real space} in the main text.
    The HF solutions at these points correspond to vortex lattice (VL), 
    Landau-level Wigner crystal (LLWC) and Landau-level-hole Wigner crystal (LLhWC) states as 
    discussed in the main text.  
    For Wigner crystal states in which interlayer coherence is absent,
    the pairing amplitudes and therefore the length of arrows are identically zero.
}
    \label{fig:more results}
\end{figure*}

At strong $B$ or at smaller values of $\Delta E$, we find model-parameter 
regions in which the Wigner crystal and vortex lattice states we find involve either many
Landau levels or order that is associated mainly with $n>0$ Landau levels. 
%As $B$ increases higher Landau levels are mixed more strongly in the vortex lattice states.
In Fig.~\ref{fig:more results}, we illustrate
the spatial patterns of the pairing amplitudes and charge densities
in Wigner crystal and vortex lattice states at several points in the 
$(\Delta E,B)$ parameter space outside the regions focused on in the main text.
Because higher Landau levels form factors play an important role in these solutions, 
the pairing and charge density patterns have more structure.
In some cases the condensate regions $A_{ex}$ between vortices shrink, 
implying that the interaction parameter $U$ in our Bose-Hubbard lattice 
should increase - increasing the strength of quantum phase fluctuations.
%for vortex lattices that involve higher Landau levels.

\end{document}